\documentclass[aps, prd, superscriptaddress, nofootinbib, preprintnumbers, showpacs, showkeys]{revtex4}

\usepackage{amsmath}
\usepackage{amssymb}
\usepackage{anysize}
\usepackage{amsbsy}
\usepackage{amsthm}
\usepackage{graphicx}
\usepackage{multirow}
\usepackage{setspace}
\usepackage[OT1,T1]{fontenc}
\usepackage{color}
\usepackage{lscape} 

\newcommand {\tr}{\mbox{tr}}
\newcommand {\Tr}{\mbox{Tr}}

\newcommand {\fhalf}{\frac{1}{2}}
\newcommand {\fquater}{\frac{1}{4}}

\newcommand{\vpsi}{\boldsymbol{\psi}}

\newcounter {definition}[section]

\newcommand {\bea}{\begin{eqnarray}}
\newcommand {\eea}{\end{eqnarray}}
\newcommand {\be}{\begin{equation}}
\newcommand {\ee}{\end{equation}}
\newcommand {\ba}{\begin{array}}
\newcommand {\ea}{\end{array}}
\newcommand {\bt}{\begin{theorem}}
\newcommand {\et}{\end{theorem}}
\newcommand {\bfig}{\begin{figure}[ht]}
\newcommand {\efig}{\end{figure}}
\newcommand {\bc}{\begin{center}}
\newcommand {\ec}{\end{center}}

\newcommand {\lsb}{\left[}
\newcommand {\rsb}{\right]}
\newcommand {\lb}{\left(}
\newcommand {\rb}{\right)}

\newcommand {\calcl}{{\cal L}}

\newcommand {\calco}{{\cal O}}

\newcommand {\longvert}{\bigg|} 

\newcommand {\refer}[1]{(\ref{#1})}

\renewcommand{\epsilon}{\varepsilon}

\newcommand {\dd}{\mathrm{d}}

\begin{document}

\setstretch{1.5}

\title{New spherically symmetric solutions in Einstein-Yang-Mills-Higgs model}

\author{Junji Jia}
\email{jjia5@uwo.ca}
\affiliation{ Department of Applied
Mathematics, University of Western Ontario, London, Ontario N6A 5B7,
Canada}
\vspace{2cm}

\begin{abstract}
We study classical solutions in the $SU(2)$
Einstein-Yang-Mills-Higgs theory. The spherically symmetric
ans\"{a}tze for all fields are given and the equations of motion are derived as a
system of ordinary differential equations. The asymptotics and the boundary conditions
at space origin for regular solutions and at event
horizon for black hole solutions are studied. Using the shooting method, we found numerical solutions to the theory. For regular
solutions, we find two new sets of asymptotically flat solutions. Each of these sets contains continua of solutions in the
parameter space spanned by the shooting parameters. The solutions
bifurcate along these parameter curves and the bifurcation are
argued to be due to the internal structure of the model. Both sets of the solutions are asymptotically flat but one is exponentially so and the other is so with oscillations.  For black
holes, a new set of boundary
conditions is studied and it is found that there also exists a continuum
of black hole solutions in parameter space and
similar bifurcation behavior is also present to these solutions. The $SU(2)$ charges of these solutions are found zero and these solutions are proven to be unstable.
\end{abstract}



\maketitle
\section{Introduction}

The first regular solution of the Einstein-Yang-Mills (EYM) theory
found by Bartnik and McKinnon \cite{rb} two decades ago stimulated
intensive research into Einstein-non-Abelian gauge theories
because of their rich geometrical and physical structure. Soon
after, the black hole solution and the violation of the no-hair
conjecture for these black holes were discovered \cite{pbizon,km}.
Because of this, this model was extended to a few other models, such
as a Einstein-Skyrme \cite{Bizon:1992gb, Heusler:1991xx}, Einstein-Yang-Mills-dilaton \cite{Lavrelashvili:1992ia}, Einstein-non-Abelian-Proca \cite{bs, Maeda:1993ap} and Einstein-Yang-Mills Higgs \cite{bs, Aichelburg:1992st} theories (see \cite{vg} for review). Black holes in these theories also violate the no-hair conjecture, and some of them are even claimed to be stable.

Among these, it is found in Ref. \cite{bs} that in the spontaneously broken phase of the EYMH model the black hole solution can possess a non-trivial field structure outside horizon. However it is not known whether there exist other solutions (in particular, black hole solutions that violate the no-hair conjecture) to this EYMH theory and whether these solutions are stable or not. In this paper, we extend the work of Ref. \cite{bs} by thoroughly studying other non-magnetically charged and asymptotically flat solutions of the EYMH theory with a Higgs doublet, a Higgs potential and a cosmological constant. Besides the solutions in Ref. \cite{bs}, we also found two other set of regular solutions and one more set of black hole solutions in this theory. One set of the regular solutions and the black hole solutions are for the minimal EYMH model (with scalar potential equal to zero) and the other set of regular solutions has an oscillatory decaying feature. The black hole solutions are characterized by the nodes of fields and therefore also provide a counter example for the no-hair conjecture. All these solutions are proven to be unstable. Therefore if one demands stability in the no-hair conjecture, the black hole solutions found here will not conflict with it.

The paper is organized as follows. In section \ref{secmodel}, we first describe the
EYMH model. Then we give the
spherically symmetric ans\"{a}tze for the metric,
Yang-Mills field and Higgs field, and derive the equations
of motion. In section \ref{secbc}, we study the boundary conditions and asymptotic behavior
that are compatible with the field equations and certain physical
criteria. In sections \ref{secrs} and \ref{secbs}, numerical
solutions for both regular and black hole boundary conditions are
found using the shooting method, and their general features are discussed. We emphasis that all solutions found here are new solutions that were not found in previous studies. In the
last section \ref{secdis}, we discuss the stability of the solutions and discuss possible extensions.

\section{Model, Equations of motion and Ans\"{a}tze}\label{secmodel}
The action of the EYMH model could be written as \be
S=\int\dd^4x\sqrt{-g}\lb\calcl_g+\calcl_m\rb\label{eqld}\ee where
$g=\det{g_{\mu\nu}}$ and the Lagrangian densities of the gravity
part $\calcl_g$ and the matter part $\calcl_m$ are
\begin{eqnarray}
\calcl_g&=&\frac{1}{16 \pi G_N}R+\Lambda,\label{laggrav}\\
\calcl_m&=&-\frac{1}{4}F_{\mu\nu}^{(a)}F_{\lambda\rho}^{(a)}
g^{\mu\lambda}g^{\nu\rho}-[D_{\mu}\Phi]^{\dag}[D_{\nu}\Phi]g^{\mu\nu}-V(\Phi^{\dag}\Phi),\\
V(\Phi^{\dag}\Phi)&=&m^2\Phi^{\dag}\Phi+\lambda(\Phi^{\dag}\Phi)^2.\end{eqnarray}
The vector field $A_{\mu}$ is of
$SU(2)$ and the doublet Higgs field $\Phi$ has two complex
components: \be
A_{\mu}=A_{\mu}^{(a)}\frac{\tau^a}{2},~~\Phi=\lb\ba{c}\phi_1\\
\phi_2\ea\rb\ee where $\tau^a/2$ $(a=1,2,3)$ are the $SU(2)$
generators and $\phi_1$ and $\phi_2$ are complex fields. The field strength
$F_{\mu\nu}=\partial_{\mu}A_{\nu}-\partial_{\nu}A_{\mu}-i\tilde{g}[A_{\mu},A_{\nu}]$
is defined as in flat spacetime and
$D_{\mu}=\partial_{\mu}-i\tilde{g}A_{\mu}$ is the covariant
derivative, where $\tilde{g}$ is the gauge coupling constant.
$R_{\mu\nu}$ and $R$ are the Ricci tensor and scalar respectively,
and $m$ and $\lambda$ are the mass and coupling constant of the
Higgs field. $\Lambda$ appears as a parameter to assure the positive
definiteness of the energy.

Variation of the action \refer{eqld} with respect to
$g^{\mu\nu}$, $A_{\nu}^{(a)}$ and $\Phi^{\dag}$ gives rise to the
following equations of motion
\begin{subequations}\label{generalodes}
\begin{eqnarray}
&&R_{\mu\nu}-\frac{1}{2}g_{\mu\nu}R-\frac{1}{2}g_{\mu\nu}\Lambda\cdot
16\pi G_N \nonumber\\
&=& 8\pi G_N\left\{ F_{\mu
\lambda}^{(a)}F_{\nu\rho}^{(a)}g^{\lambda\rho}-\frac{1}{4}g_{\mu\nu}F_{\lambda\rho}^{(a)}F_{\sigma\eta}^{(a)}g^{\lambda
\sigma}g^{\rho \eta}+2(D_{\mu}\Phi)^{\dag}(D_{\nu}\Phi)\right.\nonumber\\
&&\left.-g_{\mu\nu} (D_{\lambda}\Phi)^{\dag}(D_{\rho}\Phi)g^{\lambda\rho}-\left[m^2\Phi^{\dag}\Phi+
\lambda(\Phi^{\dag}\Phi)^2\right]g_{\mu\nu}\right\} \label{eomg},\\
&&\partial_{\mu}(\sqrt{-g}F^{\mu\nu(a)})+\sqrt{-g}\tilde{g}\epsilon_{abc}A^{(b)}_{\mu}F^{\mu\nu(c)}
-2\Tr[(\partial^{\mu}\tau^a)\tau^b]F_{\mu\nu}^{(b)}\nonumber\\
&&-\sqrt{-g}\left[i\tilde{g}\left(\Phi^{\dag}\frac{\tau^a}{2}\partial^{\nu}\Phi
-\partial^{\nu}\Phi^{\dag}\frac{\tau^a}{2}\Phi\right)+\frac{\tilde{g}^2}{2}
A^{\nu (a)}\Phi^{\dag}\Phi\right]=0\label{eoma},\\
&&D_{\mu}\left[\sqrt{-g}g^{\mu\nu}D_{\nu}\Phi
\right]-\sqrt{-g}[m^2\Phi+2\lambda(\Phi^{\dag}\Phi)\Phi]=0.\label{eomp}
\end{eqnarray}
\end{subequations}
The first equation could be written in the trace-reversed form
\begin{eqnarray}
R_{\mu\nu}&=&8\pi G_N\left\{
F_{\mu\lambda}^{(a)}F_{\nu\rho}^{(a)}g^{\lambda\rho}-\frac{1}{4}g_{\mu\nu}F_{\lambda\rho}^{(a)}F_{\sigma\eta}^{(a)}g^{\lambda
\sigma}g^{\rho \eta}\right.\nonumber \\ &&\left. +2(D_{\mu}\Phi)^{\dag}(D_{\nu}\Phi)+\left[m^2\Phi^{\dag}\Phi+\lambda(\Phi^{\dag}\Phi)^2\right]g_{\mu\nu}-2\Lambda
g_{\mu\nu}\right\}.\label{eomgrev}
\end{eqnarray}

To find useful solutions to this equation system, we concentrate on
the static and spherically symmetric ans\"{a}tze of the
metric, gauge field and Higgs field. The most general form for
static and spherically symmetric metric could be written as follows
\begin{equation} \dd s^2=-T(r)^{-2}\dd t^2+R(r)^2\dd
r^2+r^2(\dd\theta^2+\sin^2\theta
\dd\varphi^2),\label{anzmetric1}\end{equation} where if we let \be
R(r)=(1-2M(r)/r)^{-1/2} \label{rmrel}, \ee  $M(r)$ could be
interpreted as the Misner-Sharp mass within radius $r$ \cite{Misner:1964je}.

For the study of black hole solutions, it is more convenient to use
an alternative metric of the form \be \dd s^2=-\lb
1-\frac{2M(r)}{r}\rb e^{-2\delta(r)}\dd t^2+\lb
1-\frac{2M(r)}{r}\rb^{-1}\dd r^2+r^2(\dd\theta^2+\sin^2\theta
\dd\varphi^2) ,\label{anzmetric2}\ee which is obtained from Eq.
\refer{anzmetric1} by defining $\delta(r)=-\ln(R(r)/T(r))$. A
regular event horizon of this metric requires that \be
M(r_h)=\frac{r_h}{2}, ~\delta(r_h)<\infty, \ee where $r_h$ is the
horizon radius. For later purpose, we can always rescale the time
coordinate in \refer{anzmetric1} and \refer{anzmetric2} by $ \dd
\tilde{t}\equiv T^{-1}(0)\dd t$ and $\dd \tilde{t}\equiv
e^{-\delta(r_h)}\dd t$ respectively, after which we have
$\tilde{T}(0)=1$ and $\tilde{\delta}(r_h)=0$. These rescalings will
simplify the numerical calculations that will be done in the next
section. Hereafter we will drop the tilde symbols.

The most general spherically symmetric form of gauge field is given
by \cite{wt,pbizon,bs}
\begin{equation}
A=\frac{1}{\tilde{g}}\{a\tau_r\dd t+b\tau_r\dd
r+[d\tau_{\theta}-(1+c)\tau_{\varphi}]\dd\theta+[(1+c)\tau_{\theta}+d\tau_{\varphi}]\sin\theta
\dd\varphi\}, \label{sphg}\end{equation} where $a,~b,~c,~d$ are real
functions that only depend on $r$ and $t$, and
$(\tau_r,\tau_{\theta},\tau_{\varphi})$ are the Lie algebra $su(2)$
bases which satisfy
$\tr(\tau_a\tau_b)=1/2\delta_{ab},~[\tau_a,\tau_b]=i\epsilon^{abc}\tau_c
~(a,b,c=r,\theta,\varphi)$. In particular, here we choose them to be
\be
\lb\ba{c}\tau_r\\
\tau_{\theta}\\
\tau_{\varphi}\ea\rb=\fhalf\lb\ba{ccc}\sin\theta\cos\varphi&\sin\theta\sin\varphi&\cos\theta\\
\cos\theta\cos\varphi&\cos\theta\sin\varphi&-\sin\theta\\
-\sin\varphi&\cos\varphi&0\ea\rb\lb\ba{c}\tau_1\\
\tau_2\\
\tau_3\ea\rb\equiv\frac{\textbf{J}}{2}\lb\ba{c}\tau_1\\
\tau_2\\
\tau_3\ea\rb\ee where $\textbf{J}$ is just the unitary Jacobian
matrix for transformation from cartesian to spherical coordinate and
$\tau_i~(i=1,2,3)$ are the usual Pauli matrices. The gauge field
\refer{sphg} has a residual $U(1)$ gauge transformation of the full
$SU(2)$ gauge group \bea
U&=&\exp(i\beta(r,t)\tau_r)\\
A&\to& UAU^{-1}+\frac{1}{\tilde{g}} U\dd U^{-1},\eea where
$\beta(r,t)$ is an arbitrary function. This gauge transformation
could be used to put $b=0$ in Eq. \refer{sphg} identically. For the
remaining three degrees of freedom, we eliminate two by
concentrating only on the purely magnetic and static YM field, i.e.,
setting $a=d=0$ and $c=c(r)$. With this setting, the ans\"{a}tz \refer{sphg} is reduced to
\be A=\frac{1}{\tilde{g}}\lsb
-(1+c)\tau_{\varphi}\dd\theta+(1+c)\tau_{\theta}\sin\theta
\dd\varphi\rsb\label{anzgauge}. \ee

The most general form of the Higgs field could be written as \be
\Phi(x)=\frac{1}{\sqrt{2}}\lb\ba{c}\psi_2(x)+i\psi_1(x)\\
\phi(x)-i\psi_3(x)\ea\rb~,\label{anzhiggs}\ee where we can treat the
three degrees of freedom $\psi_a~(a=1,2,3)$ as those of a vector
field $\vpsi(x)$. To yield a spherically symmetric and static energy
density, we can use a more useful ans\"{a}tz by taking $\phi(x)=\phi(r),
~\vpsi(x)=\psi(r)\bf{\hat{n}}_r$ \cite{bs}. To simplify the numerical analysis that will be conducted in the following sections, we further set $\psi=0$ henceforth.

To obtain the detailed equations in terms of the fields in these ans\"{a}tze, one can substitute the ans\"{a}tze \refer{anzmetric1},
\refer{anzgauge} and $\phi(x)$ into the system \refer{generalodes}. It is found that the system of equations of motion consists of non-linear
ordinary differential equations of $w~(w(r)\equiv c(r))$, $\phi$ and $M$
\begin{subequations}\bea r\lb1-\frac{2M}{r}\rb w''&=&\lb
(m^2\phi^2+\fhalf \lambda\phi^4-2\Lambda)r^2+\fhalf (1+w)^2\phi^2+
\frac{1}{r^2}(1-w^2)^2-\frac{2M}{r}\rb w'\nonumber\\
&&+\frac{1}{4}(1+w)r\phi^2+\frac{w(w^2-1)}{r}\label{equwre},\\
r\lb1-\frac{2M}{r}\rb \phi''&=&\lb (m^2\phi^2+\fhalf
\lambda\phi^4-2\Lambda)r^2+\fhalf (1+w)^2\phi^2+
\frac{1}{r^2}(1-w^2)^2+\frac{2M}{r}-2\rb \phi'\nonumber\\
&&+\lambda r\phi^3+\lb\frac{(1+w)^2}{2r}+m^2r\rb\phi\label{equphire},\\
M'&=&\lb 1- \frac{2M}{r}\rb\lb w'^2+\frac{1}{2}(r\phi')^2\rb
+\frac{1}{2}
\frac{(1-w^2)^2}{r^2}+\frac{1}{4}\phi^2(1+w)^2\nonumber\\
&&+\lb\fhalf m^2\phi^2+\fquater\lambda\phi^4-\Lambda\rb r^2\label{equm},\eea \label{eomsub2}\end{subequations}
and for the regular solutions, of $T$
\bea r \lb 1- \frac{2M}{r}\rb \frac{T'}{T}&=&-\lb 1-
\frac{2M}{r}\rb\lb w'^2+\frac{1}{2}(r\phi')^2\rb+\frac{1}{2}
\frac{(1-w^2)^2}{r^2}
+\frac{1}{4}\phi^2(1+w)^2\nonumber\\
&&+\lb\fhalf m^2\phi^2+\fquater\lambda\phi^4-\Lambda\rb
r^2-\frac{M}{r},\label{equtre}\eea
where $'$ refers to derivatives with respect to $r$. Note in these
equations and hereafter, we set $\tilde{g}=1$ and $G_N=1/(4\pi)$,
which are equivalent to the rescaling $r\to r\sqrt{G_N}/\tilde{g}$
and $S\to g\sqrt{G_N}S$, and therefore the system on longer depends on them.

For black hole metric \refer{anzmetric2}, we replace $T(r)$ in Eq.
\refer{equm} by $T(r)=e^{\delta(r)}(1-2M(r)/r)^{1/2}$ and Eq.
\refer{equtre} for $T(r)$ is replaced by the field equation for
$\delta(r)$: \be \delta'=
-r\phi'^2-\frac{2w'^2}{r}.\label{equdelta}\ee Note that the Eqs.
\refer{eomsub2}-\refer{equdelta} are identical to equations
(4.19)-(4.24) in Ref. \cite{bs} and equations (17)-(22) in Ref.
\cite{jj}. However here we will solve these equations by setting parameters differently.
If $\phi(r)$ is put to zero, then equations
(3)-(5) in Bartnik and Mckinnon's work \cite{rb} could be recovered.
Because the system has mirror symmetry
$\phi(r)\to-\phi(r)$, we only need to concentrate on one case (we
pick $\phi(r\to0)\leq0$) for the specification of the boundary
conditions, which are necessary in order to solve the equations
numerically.

\section{Boundary conditions}\label{secbc}
The system of equations \refer{eomsub2} and \refer{equtre} and the
system \refer{eomsub2} and \refer{equdelta} are only solvable
numerically provided proper boundary conditions are given. The
former system has singular points at $r=0$ and $r= \infty$ and the
latter has singular points at $r=r_h$ and $r=\infty$. In the following two subsections we discuss these boundary conditions.
We first examine the behavior of regular solutions at the singular
points.

\subsection{Boundary conditions for regular solutions}
As $r\to\infty$, we require that the gauge field $w(r)$ and
scalar field $\phi(r)$ approach constant asymptotic values while the
metric function $T(r)$ goes to a constant but nonzero value.

For
asymptotically flat spacetime, we can do a careful asymptotic analysis by substituting the following expansions
\begin{subequations} \bea
w(r)&\to&-1+\delta w(r),\\
\phi(r)&\to&\phi_0+\delta \phi(r),\\
M(r)&\to&M_0+\delta M(r),\\
T(r)&\to&T_0+\delta T(r),
\eea
\end{subequations} into the Eqs. \refer{eomsub2} and \refer{equtre} and treating $\delta w,~\delta \phi,~\delta M,~\delta T$ as perturbations. Here we know that the asymptotic value for $w(r)$ is -1 and $\phi_0,~M_0$ and $T_0$ refer to some asymptotic constants.

It is found that there are three possible asymptotics. The first takes the form
\begin{subequations} \bea
w(r)&\sim&-1+ce^{\phi_0r/2},\\
\phi(r)&\sim&\phi_0+\frac{1}{2}\frac{c^2e^{\phi_0r}}{\phi_0r^2},\\
M(r)&\sim&M_0+\frac{1}{2}c^2\phi_0e^{\phi_0r},\\
T(r)&\sim&T_0+\frac{T_0M_0}{r},
\eea\label{regasy1}
\end{subequations} where $c$ is some positive constant, with the following conditions on the parameters in the model
\be
\lambda=0,~\phi_0\neq0,~ m^2=\Lambda=0\label{infsys2}.\ee
The above parameter setting means that the scalar potential is zero and there is only a kinetic energy of the scalar field that is coupled to gravity. Therefore this case corresponds to the minimal EYMH theory.

The second is given by \footnote{The function $p(r)$ in \refer{phiasy2} and \refer{masy2} is given by a complicated differential equation that can not be solved analytically. A numerical inspection indicates that it is a function that increases slower than $r^1$. The asymptotics for $M(r)$ is obtained by setting $c_1=0$ for $\phi(r)$. The general form for $M(r\to\infty)$ without setting $c_1$ (or $c_2$) to zero is given by an equation that dose not allow us to get a compact solution. However, the oscillatory feature for $M(r)$ in this case should still be present. }
\begin{subequations} \bea
w(r)&\sim&-1+\frac{c}{r},\\
\phi(r)&\sim&\frac{1}{rp(r)}\lb c_1\sin\lb \sqrt{-m^2}r\rb+c_2\cos\lb\sqrt{-m^2}r\rb\rb,\label{phiasy2}\\
M(r)&\sim&M_0+\frac{1}{2}c_2^2\int \dd r\frac{1}{p(r)^2}\cos\lb2\sqrt{-m^2}r\rb,\label{masy2}\\
T(r)&\sim&T_0+\frac{T_0M_0}{r},
\eea \label{regasy2}
\end{subequations} where $c$ and $c_1,~c_2$ are some constants and $p(r\to\infty)>1$ is a slowly increasing function (slower than $r^1$), with the condition $\Lambda=0$.

The last is the asymptotics that satisfy
\be \phi_0^2=\frac{-m^2}{\lambda},~\Lambda=-\fquater\frac{m^4}{\lambda}, ~\lambda\neq0\label{infsys1}.\ee The equation system with the last asymptotics has been
solved in Ref. \cite{bs}. Therefore we will only concentrate on the first two cases. We emphasis that the above three cases are all the allowed asymptotics for the EYMH theory under the demand that the solutions are asymptotically flat.

For $r=0$, regularity of the metric and fields requires that there
exists a series solution to each function and in particular
$T(0)\neq0$. We also impose the finiteness condition at $r=0$ for
the energy density $T_{00}$, which could be calculated from the
right side of Eq. \refer{eomgrev}. Using the field equation
\refer{equtre}, one can show that $T_{00}$ satisfies \be
T_{00}(r\to 0)\sim\frac{M'(r)}{T(r)^2r^2} \label{energydensity}\ee and thus the
finiteness of $T_{00}(0)$ requires $M(r)\sim \calco(r^3)$ as
$r\to0$. These conditions directly lead to the following two
possible sets of boundary conditions at small $r$, classified
according to the number of nodes, denoted by $k$ ($k\in
\mathbb{Z}^+$), of $w(r)$. For odd-$k$ solution, we have
\begin{subequations}\bea w(r)&=&1+ar^2+w_4r^4
+\calco(r^6)\label{reg1wr}\\
\phi(r)&=&b_1r+\phi_3r^3+\calco(r^5)\label{phiexpen}\\
M(r)&=&M_3r^3+\calco(r^5)\\
T(r)&=&1+T_2r^2+\calco(r^4),\eea where
\bea w_4&=&\frac{1}{20}\lb 16a^3+6a^2+8a(b_1^2-\Lambda)+b_1^2\rb\label{reg1wc}\\
\phi_3&=&\lsb\frac{1}{10}\lb 3b_1^2+m^2+8a^2+2a\rb-\frac{4}{15}\Lambda\rsb b_1\\
M_3&=&\frac{1}{2}b_1^2+2a^2-\frac{1}{3}\Lambda\\
T_2&=&-2a^2-\frac{1}{3}\Lambda\label{reg1p4}.\eea
\label{reg1}\end{subequations} For even $k$
solutions, the boundary condition is
\begin{subequations}
\bea w(r)&=&-1+ar^2+w_4r^4+\calco(r^6)\label{reg2wr}\\
\phi(r)&=&b_0+\phi_2r^2+\calco(r^4)\\
M(r)&=&M_3r^3+\calco(r^5)\\
T(r)&=&1+T_2 r^2+\calco(r^4),\eea where \bea
w_4&=&\frac{4}{5}a^3-\frac{3}{10}a^2+\frac{1}{40}\lb 1+4\lambda
b_0^2+8m^2\rb b_0^2a -\frac{2}{5}\Lambda a \label{reg2wc}\\
\phi_2&=&\frac{1}{6}\lb \lambda b_0^2+m^2\rb b_0 \\
M_3&=&2a^2+\frac{1}{12}\lb 2m^2+\lambda b_0^2\rb b_0^2-\frac{1}{3}\Lambda\\
T_2&=&-2a^2+\frac{1}{12}\lb 2m^2+\lambda b_0^2\rb
b_0^2-\frac{1}{3}\Lambda
\label{reg2p4}.\eea\label{reg2}\end{subequations}

\subsection{Boundary conditions for black hole solutions}
For black hole solutions, we have to set  \be
M(r_h)=\frac{r_h}{2},~\delta(r_h)=0\ee and require the gauge and
Higgs field to be finite and smooth at $r=r_h$ in order to have a
regular event horizon. Using Taylor expansions for fields $w(r)$, $\phi(r)$ and $M(r)$ near the horizon, we can get a valid set of boundary
data at $r=r_h$: \begin{subequations} \label{bhbc}\bea
w'(r_h)&=& \frac{(\phi_h/2)^2(1+w_h) r_h^2-(1-w_h^2) w_h}{r_h-(
1-w_h^2)^2/r_h-2\lb\phi_h/2\rb^2(1+w_h)^2r_h-(m^2\phi_h^2+\lambda\phi^4/2-2\Lambda)r_h^3},\\
\phi'(r_h)&=& \frac{(1+w_h)^2\phi_h/2+
(m^2+\lambda\phi_h^2)\phi_hr_h^2}{r_h-(
1-w_h^2)^2/r_h-2\lb\phi_h/2\rb^2(1+w_h)^2r_h-(m^2\phi_h^2+\lambda\phi^4/2-2\Lambda)r_h^3},\\
M'(r_h)&=&\frac{1}{2}\lb
m^2\phi_h^2+\fhalf\lambda\phi_h^4-2\Lambda\rb
r_h^2+\frac{1}{4}(1+w_h)^2\phi_h^2+\fhalf (1-w_h)^2/r_h^2, \eea
\end{subequations} where $w(r_h)\equiv w_h$ and $\phi(r_h)\equiv
\phi_h$. At $r=\infty$, the same asymptotic analysis as in the previous subsection can be done. It is found that for the case \refer{regasy1}, $\delta (r)$ takes the form
\be \delta(r)\sim \delta_0-\frac{1}{2}\frac{c^2\phi_0e^{\phi_0r}}{r},\label{bhasy1}\ee  and for the case \refer{regasy2}
\be \delta(r)\sim \delta_0-\int\dd r ~r(\delta\phi(r)^\prime)^2, \label{bhasy2}\ee where $\delta\phi(r)$ is given in \refer{phiasy2}.

\section{Regular Solutions}\label{secrs}

With conditions \refer{reg1} or \refer{reg2} at $r=0$, we use the
shooting method to solve the Eqs. \refer{eomsub2} and \refer{equtre}
numerically. Using a standard ordinary differential equation solver,
we evaluate the initial data for the functions at $r=10^{-2}$ and
use tolerance $10^{-12}$ to shoot the parameters $(a,~b_1)$ for
conditions \refer{reg1} and $(a,~b_0)$ for conditions \refer{reg2}
and integrate towards $r=\infty$ to match asymptotics \refer{regasy1}\refer{regasy2}. A drawback of
the two parameter shooting procedure is its slow convergence.
Therefore, we will limit our study to only $k=1$ and/or $k=2$
solutions. Scalar mass $m$, scalar
coupling $\lambda$ and cosmological constant $\Lambda$ are the three
parameters on which the shooting process depends. We will clarify
how they affect the existence and features of the solution. In the
following, subsection \ref{subsecinf2} contains solutions that match
asymptotics \refer{regasy1}, while in subsection \ref{subsecinf3},
solutions match condition \refer{regasy2} are shown.

\subsection{Asymptotically flat solutions with asymptotics \refer{regasy1}\label{subsecinf2}}
\begin{figure}[htp!]
\includegraphics[scale=0.4]{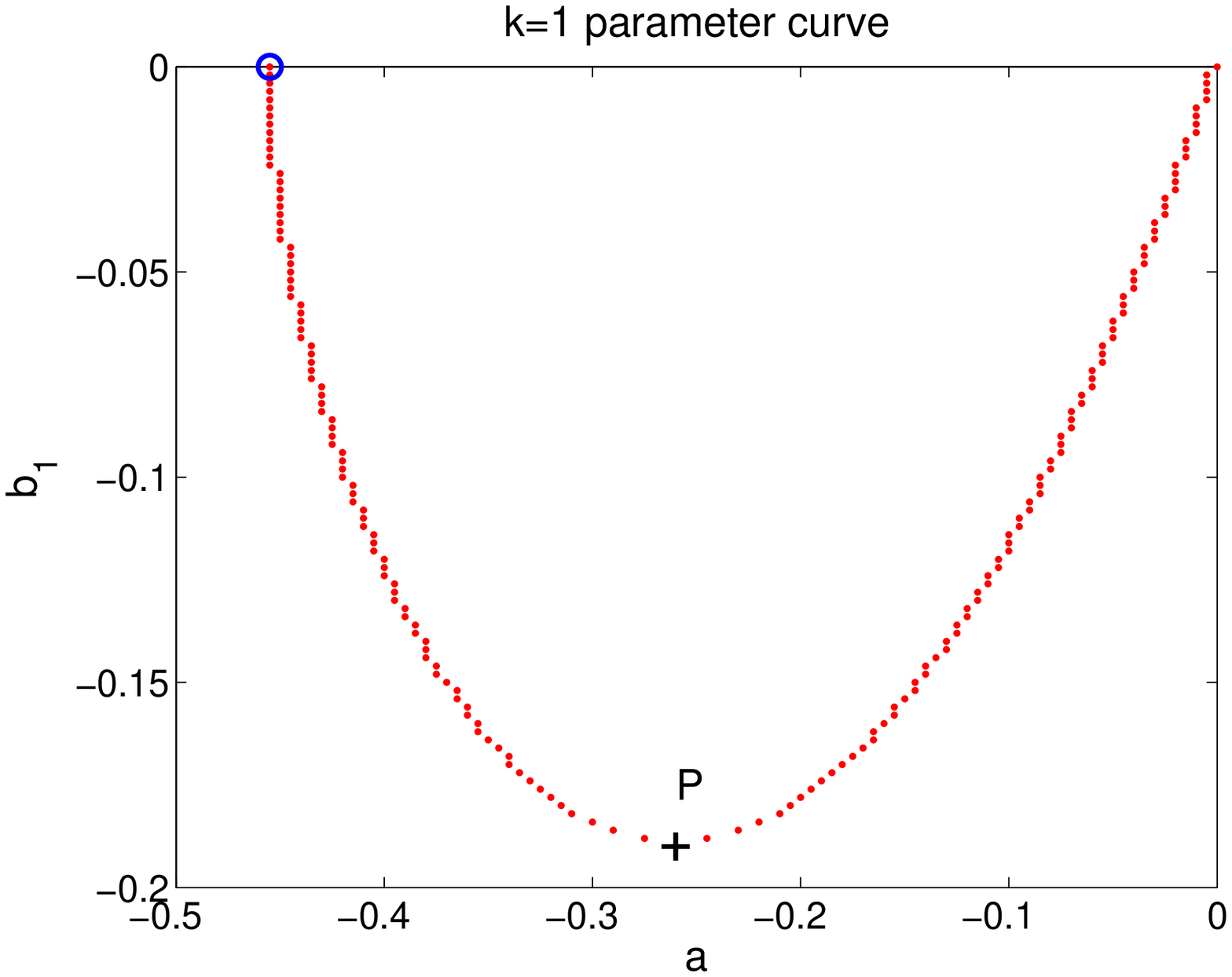}
\includegraphics[scale=0.4]{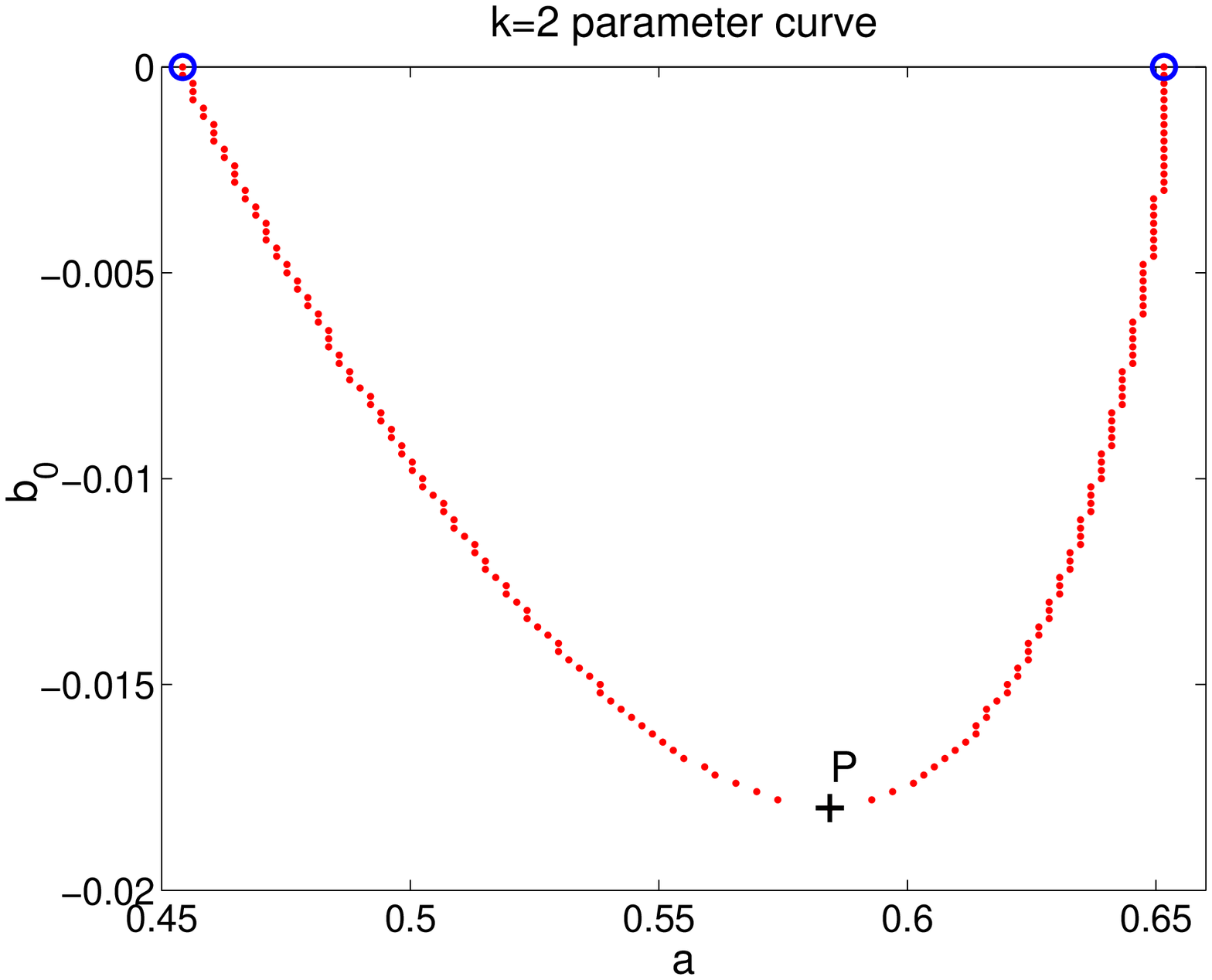}
\caption{\label{reg1curves} The $k=1$ and $k=2$ parameter curves.
Each point on these curves corresponds to a $k$ node solution. The
circle(s) at $b_1=0$ is the $a_1$ and at $b_0=0$ are $a_1$ and $a_2$
of EYM theory.}
\end{figure}

For system with both boundary conditions \refer{reg1} and
\refer{reg2} at $r=0$ and \refer{regasy1} at $r=\infty$, we found
that when $b_{1,0}$($b_1$ or $b_0$)=0, there exist only solutions
with $a$ taking discrete but infinitely many values whose magnitude
falls in the range of $0.453$ and $0.707$. We denote these values by
$a_k$, where $k$ are positive integers that equal the numbers of
nodes of $w(r)$. Indeed these solutions are just the solutions of
EYM theory discovered in Ref. \cite{rb} and a list of values of
$a_k$ and the position of the nodes of $w(r)$ can be found in Ref.
\cite{km}. In these solutions, $\phi(r)$ is identically zero and
this identically vanishing $\phi(r)$ is a good solution because it
dose not contribute any kinetic or potential energy to the system
and therefore allows the possibility of asymptotic flatness.

\begin{figure}[htp!]
\includegraphics[scale=0.4]{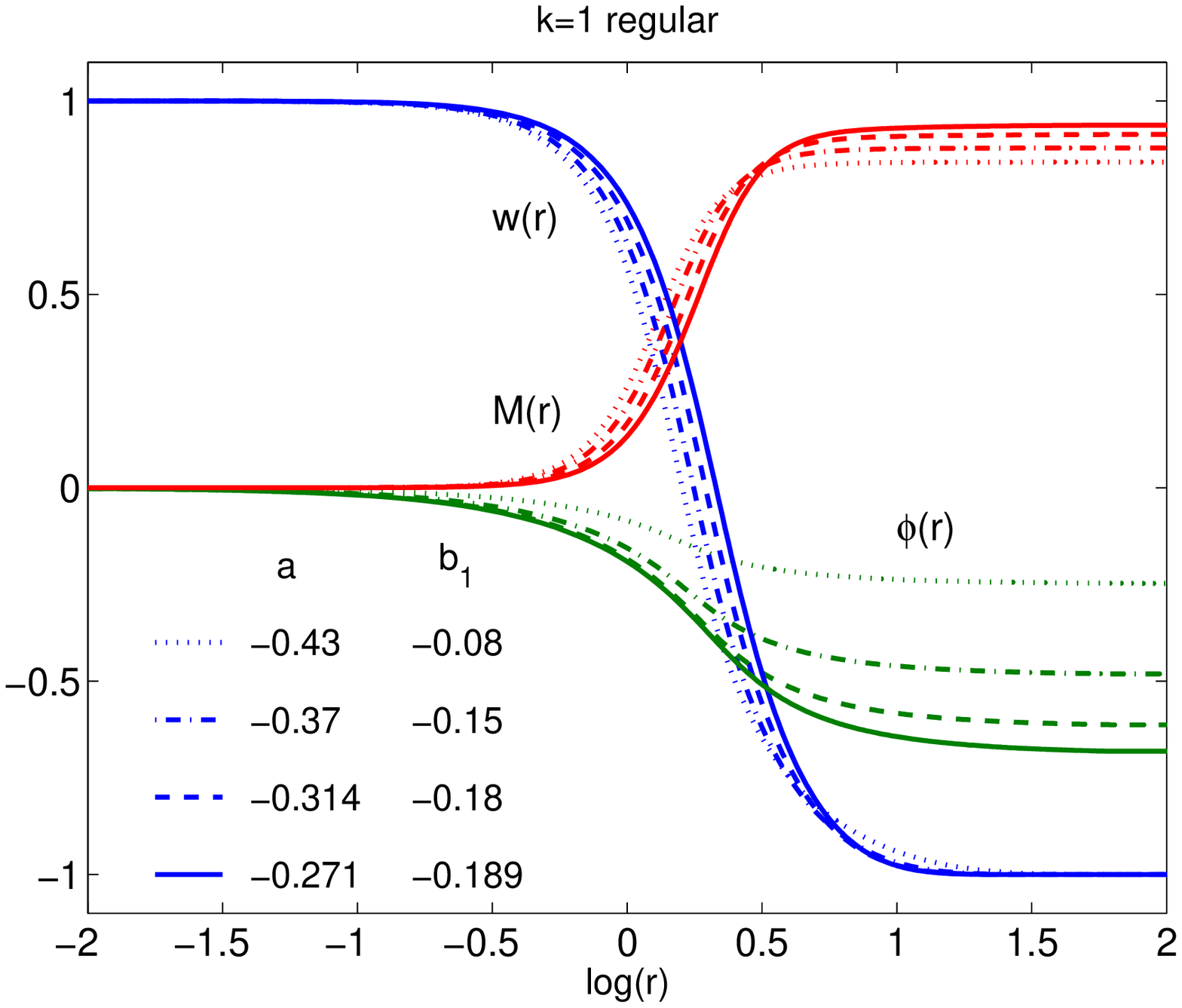}
\includegraphics[scale=0.4]{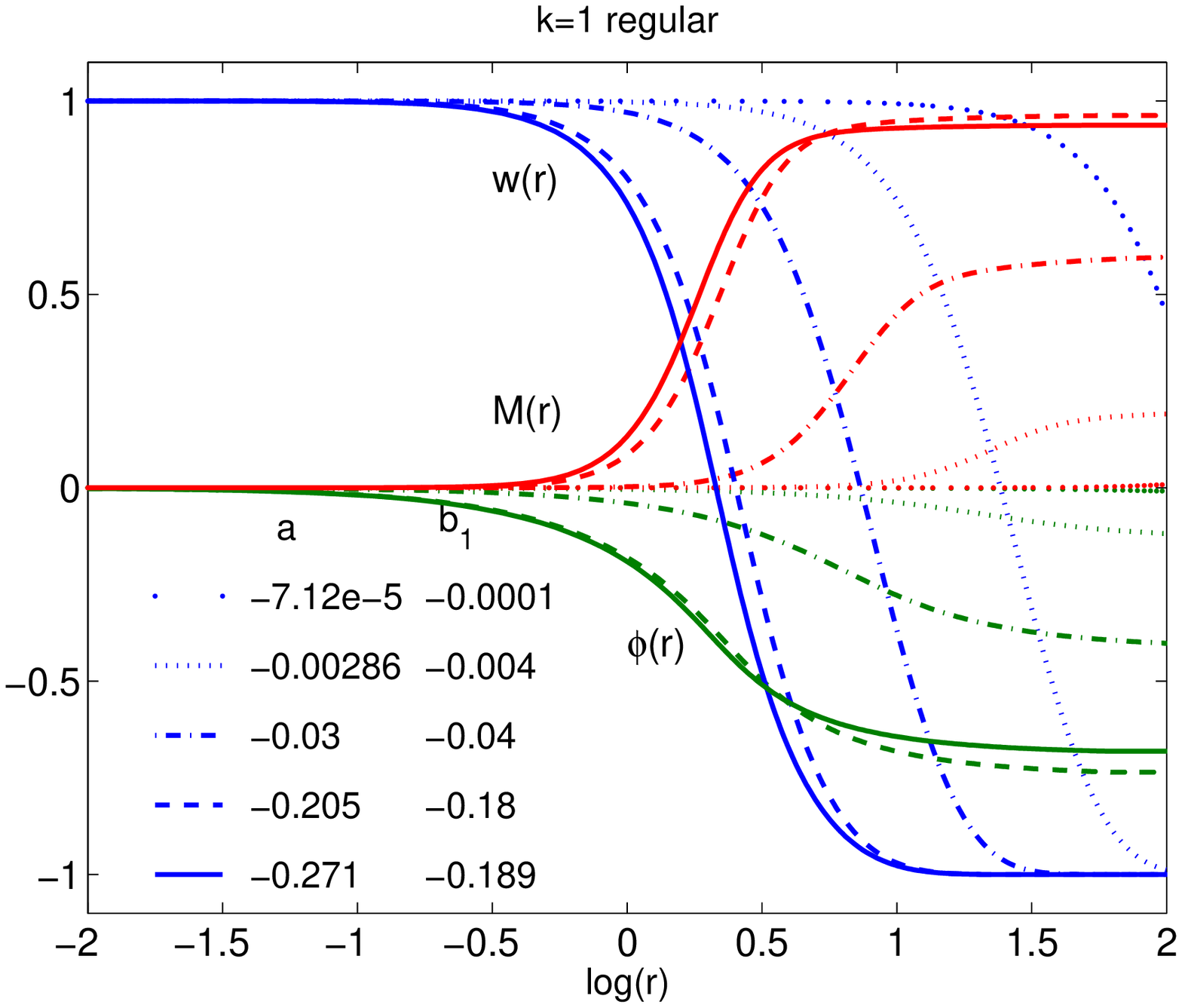}\\
\includegraphics[scale=0.4]{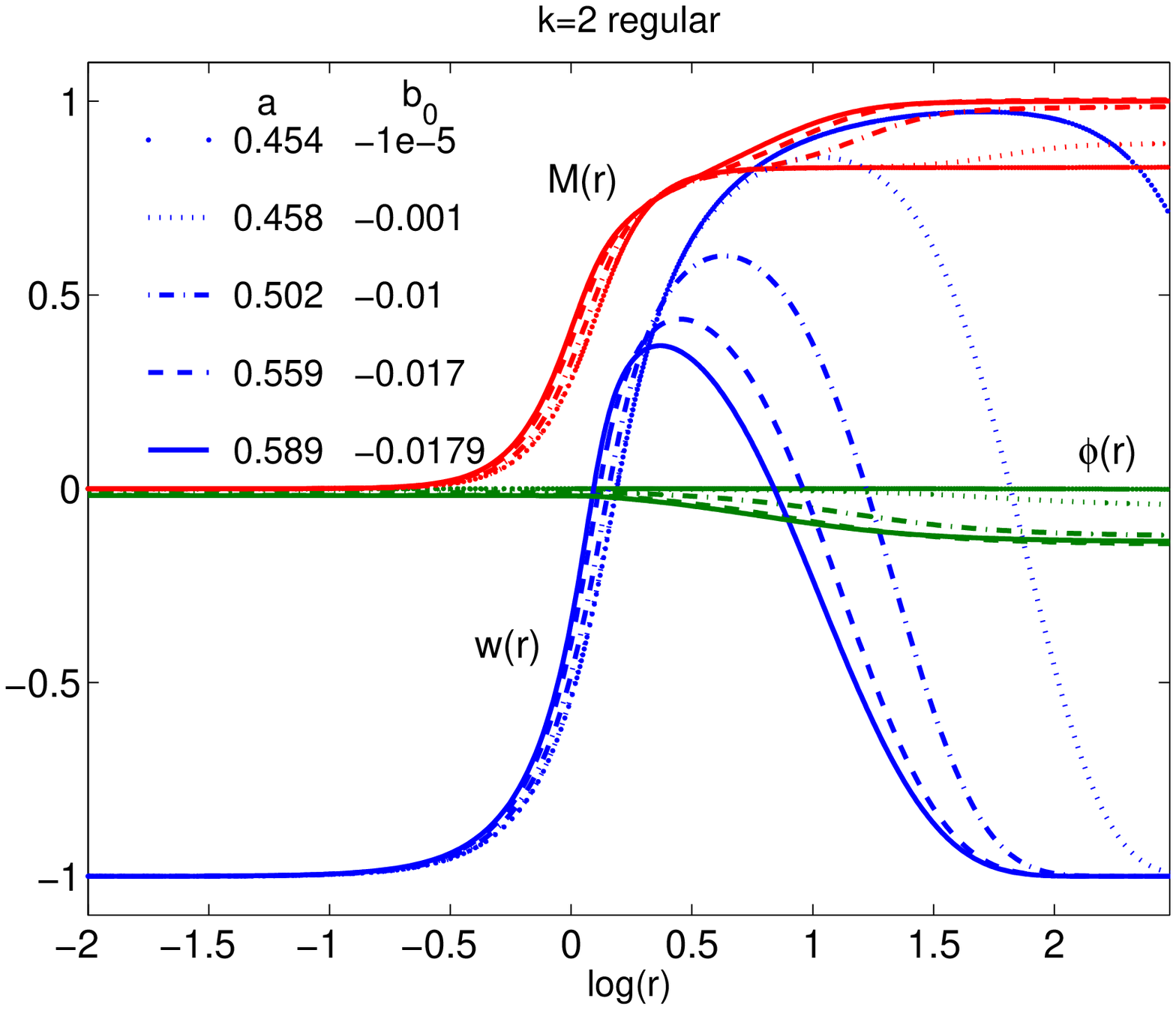}
\includegraphics[scale=0.4]{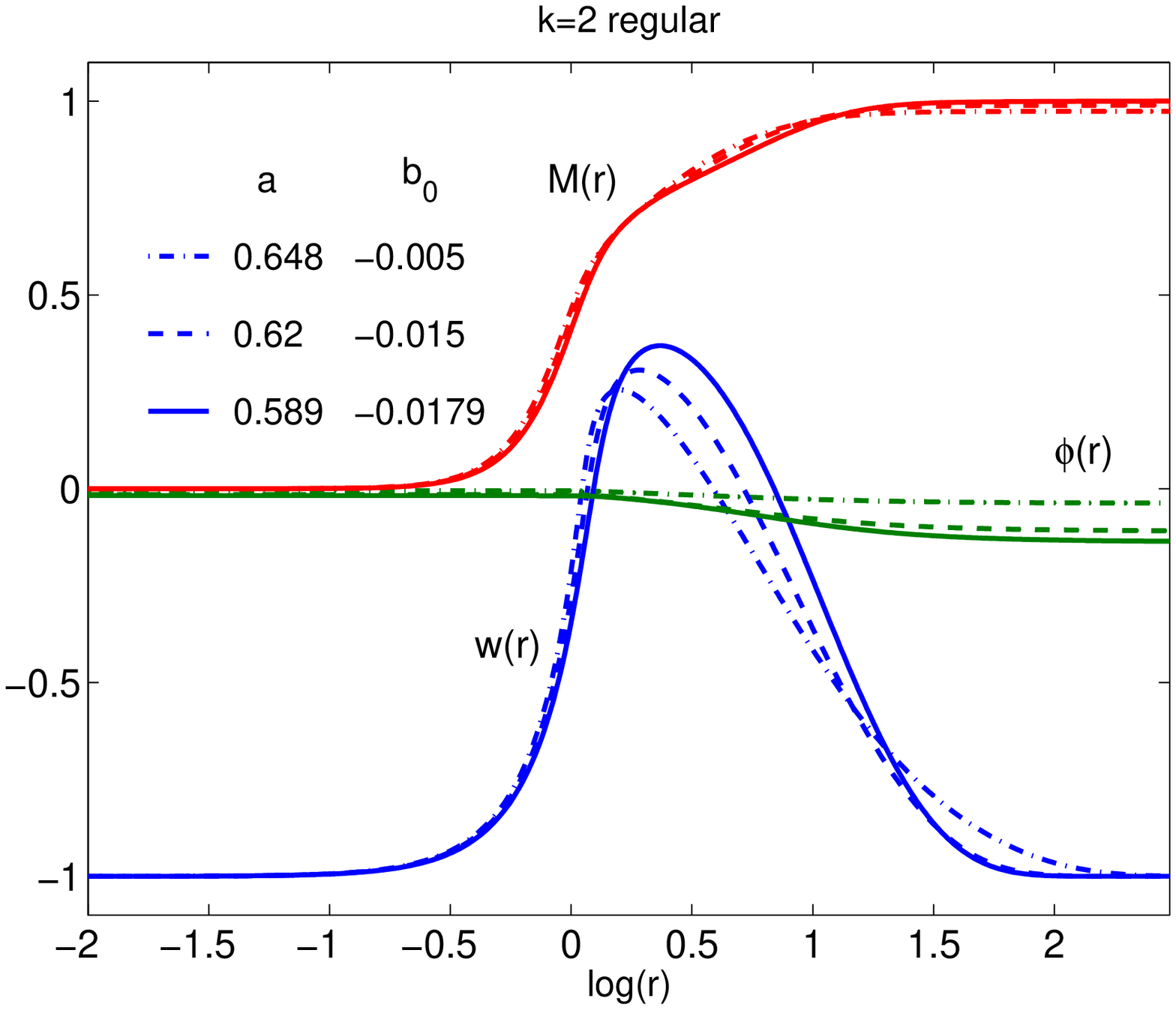}
\caption{\label{reg1sols} The $k=1$ and $k=2$ regular solutions
bifurcating at $P$ of the parameter curve. The solid curves are
solutions corresponding to peaks $P$ in the parameter curves.}
\end{figure}

The solutions of our EYMH theory \refer{eqld} and that are
unique to boundary conditions \refer{reg1} or \refer{reg2} are those
with $b_{1,0}\neq0$. For each positive integer $k$, we found that
there exists in the shooting parameter space spanned by $a$ and
$b_{1,0}$ a continuous curve, each point of which can give a $k$
node solution. The two end points of this curve are the
$(a_k,b_{1,0}=0)$ and $(a_{k-1},b_{1,0}=0)$, while the points in
between have $a_k<a<a_{k-1}$ and $b_{1,0}\neq0$ (for $k=1,2$ curves,
see Fig. \ref{reg1curves}).

The main feature of these solutions is that there is a bifurcation
of the solutions, with bifurcation points being the peak points
(points $P$ in Fig. \ref{reg1curves}) of the parameter curves. On
each parameter curve, from the peaks along the negative
direction of $|a|$ to point $(a_{k-1},b_{1,0}=0)$, one find that the
configuration of the solutions resembles that of the $k-1$ node EYM
solution; from the peaks to $(a_k,b_{1,0}=0)$, the
solutions resemble the $k$ node EYM solution (see Fig.
\ref{reg1sols}); while the solutions at the peaks $P$ have the
feature that their asymptotic mass (or energy) is relatively larger
than other solutions of the same parameter curve. This variation of
the solutions along the parameter curve manifestly shows that the
solutions to the EYMH theory naturally cover that of the EYM
solution, as one should expect.

This bifurcation is very different both superficially and in its
origin from the bifurcation behavior discovered in Ref. \cite{bs}.
There the bifurcation depends on the variation of one original
parameter in the theory -- the mass of scalar, while here the
bifurcation occurs purely in the shooting parameter space. For the
origin of the phenomenon, in Ref. \cite{bs} the EYHM theory is not
minimally coupled: they have scalar mass and $\phi^4$ potential
terms, and the bifurcation was explained using two length scales
$L_1=\tilde{g}^{-1}G^{1/2}$ and
$L_2=\tilde{g}^{-1}\sqrt{\lambda/m^2}$. However here in this
subsection, the Higgs field is minimally coupled to the gauge field
and gravity and therefore essentially there is only one length scale
$L_1$. Therefore we can not use the same length scale argument and
we tend to believe that this bifurcation is simply due to the
variation of Higgs field $\phi(r)$ and gauge field $w(r)$ along each
parameter curve, which is purely an intrinsic structural feature of
the EYMH theory under consideration.

This is supported by the observation from equation \refer{equm}
(with $\mu=m=\lambda=\Lambda=0$) that the mass or energy of the
system is crucially dependant on the derivatives $w(r)'$, $\phi(r)'$
and the deviation of $w(r)$ from $\pm1$ and $\phi(r)$ from 0. The
derivatives represent the kinetic energy from the gauge and scalar
field; and the other terms come from the covariant derivatives of
the scalar field, where the non-Abelian field couples to itself and
the scalar field. (We emphasis that this feature is unique to
non-Abelian gauge theories.) To see how the parameters $a$ and
$b_{1,0}$ determine the solutions in Fig. \ref{reg1sols}, we take
$k=1$ for example. Along the directions of the parameter curve that
$|b|$ increase, from \refer{reg1p4} one see $\phi(0)'$ increases
while $w(0)'$ dose not change. This effectively increase the rate
that $M(r)$ grows at small $r$ and eventually leads to larger
asymptotic mass at the peak $P$ of the parameter curve. The process
that the fields and mass function evolve to larger values of $r$ has
to be determined from the field equations
\refer{eomsub2}-\refer{equtre}, whose complicated structure blocks
us from gaining further insights about the physics.

Note that throughout this subsection, the solutions found are
the solutions to the simplest EYMH models in the sense that both
scalar-gravity and scalar non-Abelian gauge field are minimally
coupled.

\subsection{Asymptotically flat solutions with asymptotics \refer{regasy2}\label{subsecinf3}}

From the asymptotics \refer{phiasy2} and \refer{masy2}, it is seen that that the asymptotic solutions to $\phi(r)$ and $M(r)$ will be oscillatory with a decreasing magnitude, provided that $m^2<0$. The situation for $m^2>0$ will produce hyperbolic asymptotics (from \refer{phiasy2} and \refer{masy2}), which corresponds to a spacetime that is far from being flat and therefore is not of interest here. Therefore in this subsection, we will always study the case with $m^2<0$. In solving the equation system with asymptotics \refer{regasy2}, we will simply set $\lambda=\frac{1}{8}$ in order to reduce the amount of calculation.

\begin{figure}[htp!]
\includegraphics[scale=0.8]{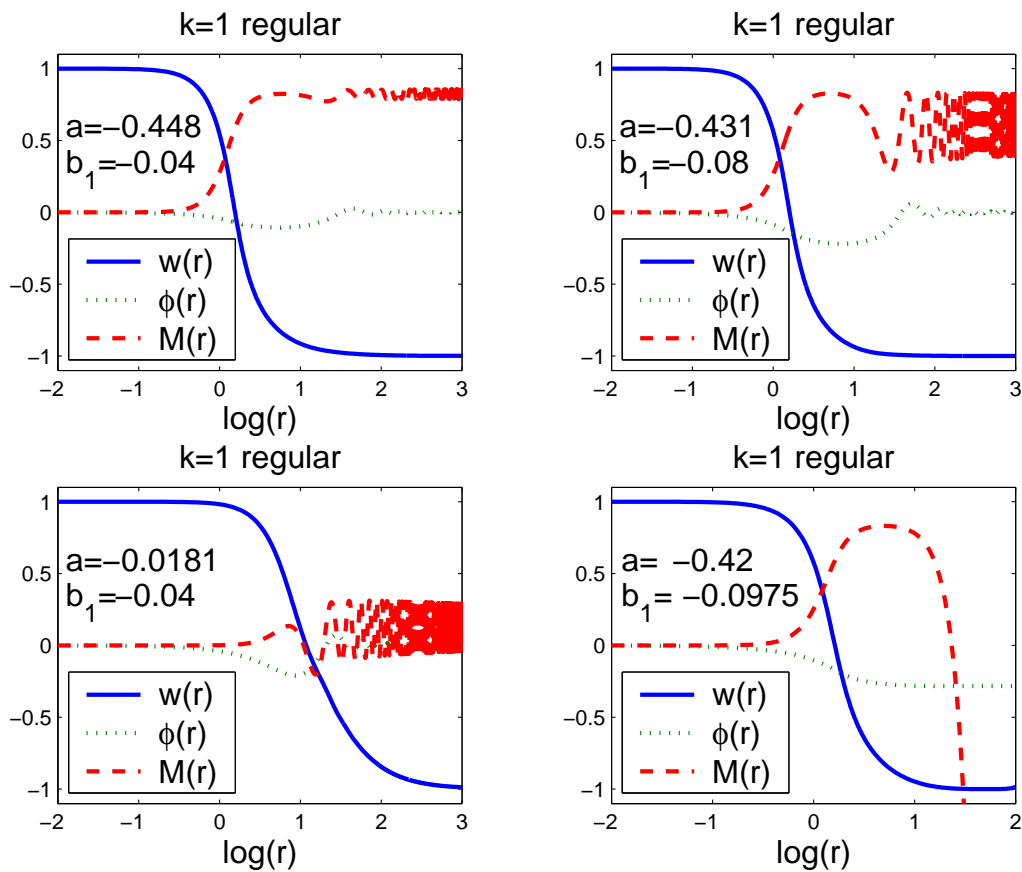}
\caption{\label{infsys3figs1} The solutions for $k=1$ to the system
\refer{eomsub2} and \refer{equtre} with asymptotics
\refer{regasy2}. For all these solutions,
$\lambda=1/8$, $\mu=0$ and $\Lambda=0$. For the lower left solution, $m^2=-0.04$. For all other solutions, $m^2=-0.01$. }
\end{figure}

\begin{figure}[htp!]
\includegraphics[scale=0.5]{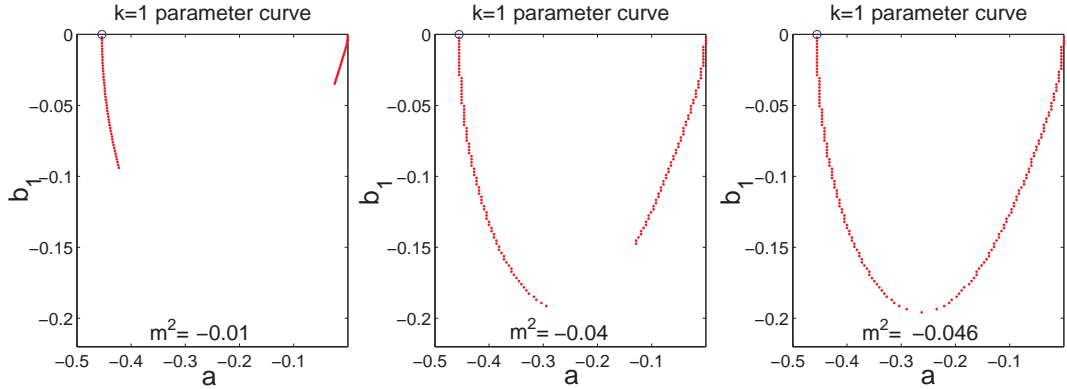}
\caption{\label{infsys3paracurve} The $k=1$ parameter curves of the
system \refer{eomsub2} and \refer{equtre} with asymptotics
\refer{regasy2} for different $m^2$'s. The two lower end points of
the left figure are $(a=-0.4228,b_1=-0.0975)$ and $(a=-0.02438,
b_1=-0.03467)$ and center figure $(a=-0.2944, b_1=-0.1914)$ and
$(a=-0.1288,b_1=-0.1474)$. }
\end{figure}

\begin{figure}[htp!]
\includegraphics[scale=0.8]{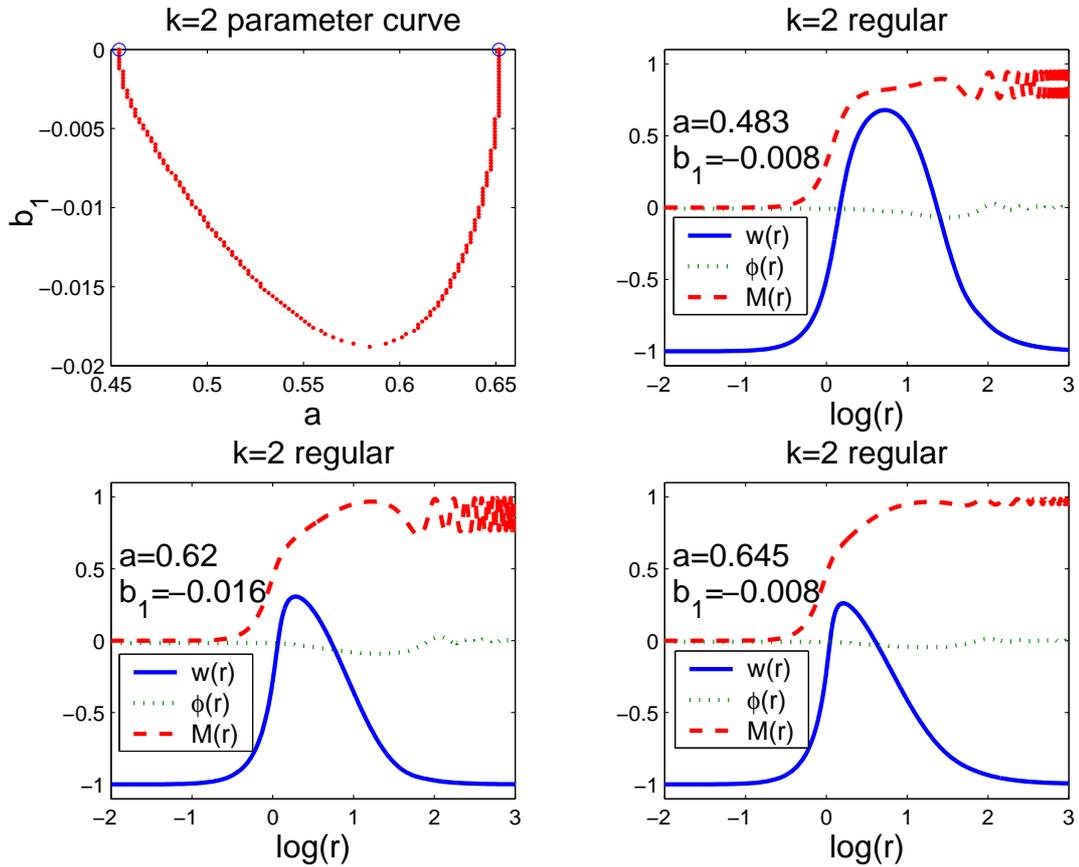}
\caption{\label{infsys3figs2} The parameter curve and solutions for
$k=2$ to the system with asymptotics \refer{regasy2}. For
all these solutions, $m^2=-0.002$, $\lambda=1/8$, $\mu=0$ and
$\Lambda=0$. }
\end{figure}

We study the solutions satisfying \refer{reg1} with $k=1$ first. When
$m^2$ is less than zero, it is found that two continuous curves in the
parameter space spanned by $a$ and $b_1$ start to emerge form the
point $(a=a_1,b_1=0)$ and $(a=0,b_1=0)$. These curves are extended as
$m^2$ decreases and finally join each other at $m^2\approx -0.046$.
Each point on these curves can give a valid solution that satisfies
the boundary conditions. These solutions in general are oscillatory
and only approach their expectation values at infinity. However, as
we decrease $b_1$ (increase $|b_1|$; see Fig.
\ref{infsys3paracurve}) along these two solution curves to their end
points, the starting radius of the oscillation of $\phi(r)$ becomes
delayed and the first minimum of $\phi(r)$ is enlarged, and
eventually at the end point of each parameter curve, $\phi(r)$
becomes flat right after it reached its nonzero minimum value. This
violates the asymptotics \refer{regasy2} and will
force $M(r)$ to diverge to negative infinity and therefore the end points do not
correspond to physical solutions (see Fig.
\ref{infsys3paracurve} lower right). For $M(r)$, with $b_1$
decreases ($|b_1|$ increases), its oscillation amplitude grows
larger until $b_1$ reaches its end point value, where the
oscillation amplitude of $M(r)$ becomes infinity. Fig.
\ref{infsys3figs1} shows the solutions for $m^2=-0.01$ and various
values $(a,b_1)$'s. Fig. \ref{infsys3paracurve} shows the extension and
joining of the parameter curves with respect to the decrease of $m^2$.

The phenomena where the parameter curves are extended with larger
$|m^2|$ and where the solutions are oscillatory could be understood in a
heuristic manner. The key is still the field equation \refer{equm}
of the mass (or energy) function. The existence of asymptotically
flat solutions for $M(r)$ depends on the balance of the kinetic term
$\fhalf (r\phi')^2$ and mass potential term $\fhalf m^2\phi^2$. As
$m^2(<0)$ decreases, the mass term provides a larger negative value
so that the allowed $\phi'$ could also be extended to a larger
value, which means $b_1$ is extended noticing Equ. \refer{phiexpen}.
While for the oscillation of $M(r)$ and the increase of the
oscillation amplitude, it purely comes form the mass term $\fhalf
m^2\phi^2$ because only this term is negative in $M(r)'$ and larger
$|m^2|$ provides larger slope for oscillation of $M(r)$. These are
confirmed by the simultaneity of oscillations of $M(r)$ and
$\phi(r)$ in the solution shown in Fig. \ref{infsys3figs1}. The
explanation of the divergence of $M(r)$ for solutions corresponding
to the lower ends of the parameter curves (left and center of Fig.
\ref{infsys3paracurve}) relies not only on Eq. \refer{equm} but
\refer{equphire}. From the latter we see that if $\phi(r)$ becomes
flat at some large $r$, one has to have $\lambda\phi^2=-m^2$, that
is, $\phi=\sqrt{-m^2/\lambda}$ and this means that there is a
negative energy in \refer{equm} $M(r>>1)'=-m^4/(4\lambda)r^2$, which
directly lead to the divergence of $M$ observed in Fig.
\ref{infsys3figs1} (lower right). The bifurcation of the solutions
for each $m^2<-0.046$ is similar to that of the solutions discussed
in the previous subsection. Again, because the bifurcation is for
different choice of the inner parameters $a$ and $b_{0,1}$ but not
with respect to different choice of $m^2$, we can not explain
the bifurcation by appealing to the existence of the two scales, as
was done in \cite{bs}.

For solutions with even nodes (see Fig. \ref{infsys3figs2}), we similarly found that when $0>m^2>-0.002$, there exist two curves in parameter space spanned by
$a$ and $b_0$ and eventually these two curves join at $m^2\approx
-0.002$. These solutions also have oscillatory $\phi(r)$ and $M(r)$
and their dependence on various parameters are quite similar as that
of $k=1$ node solutions.

\section{Black Hole Solution}\label{secbs}

\begin{figure}[htp!]
\includegraphics[scale=0.4]{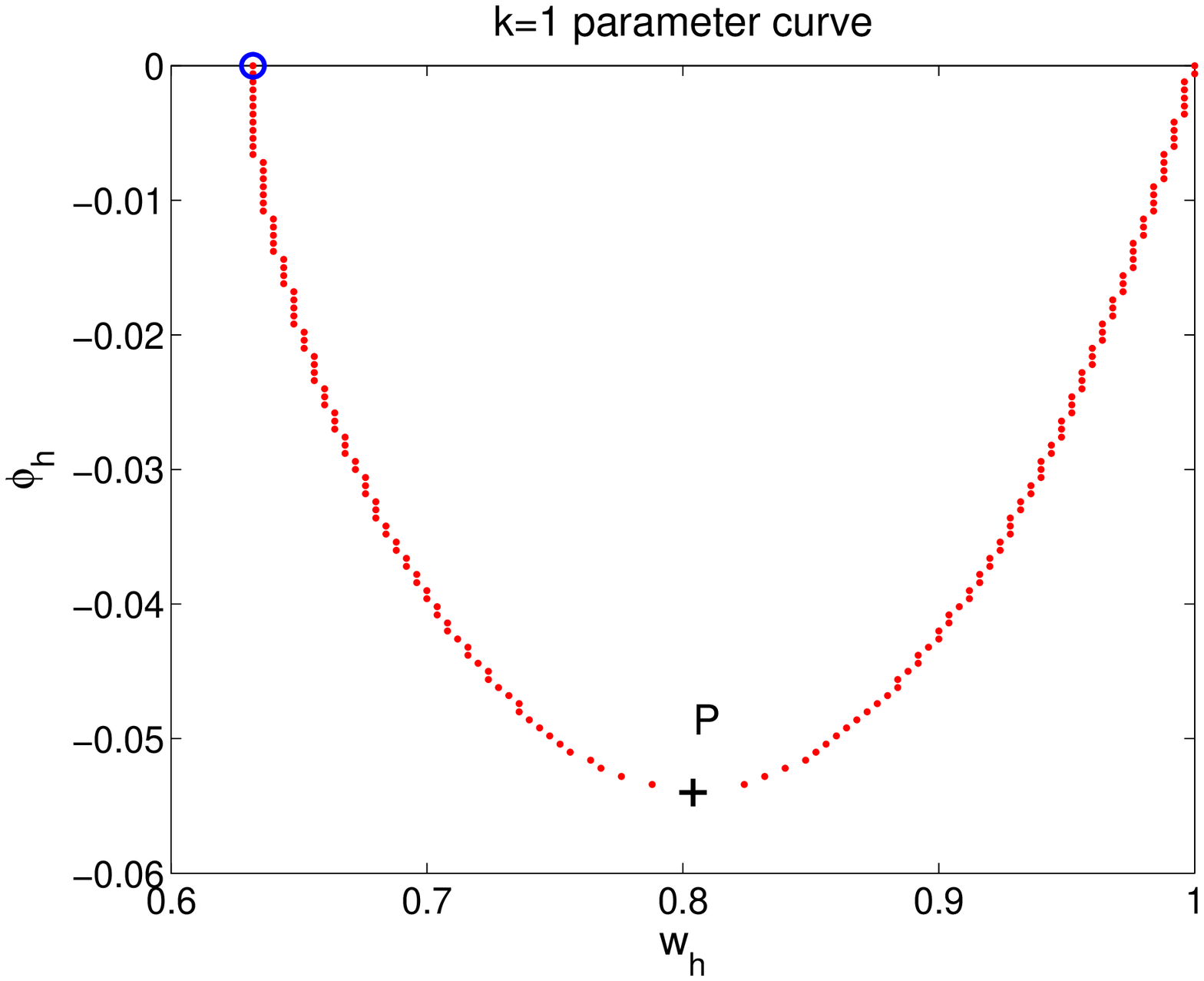}
\includegraphics[scale=0.4]{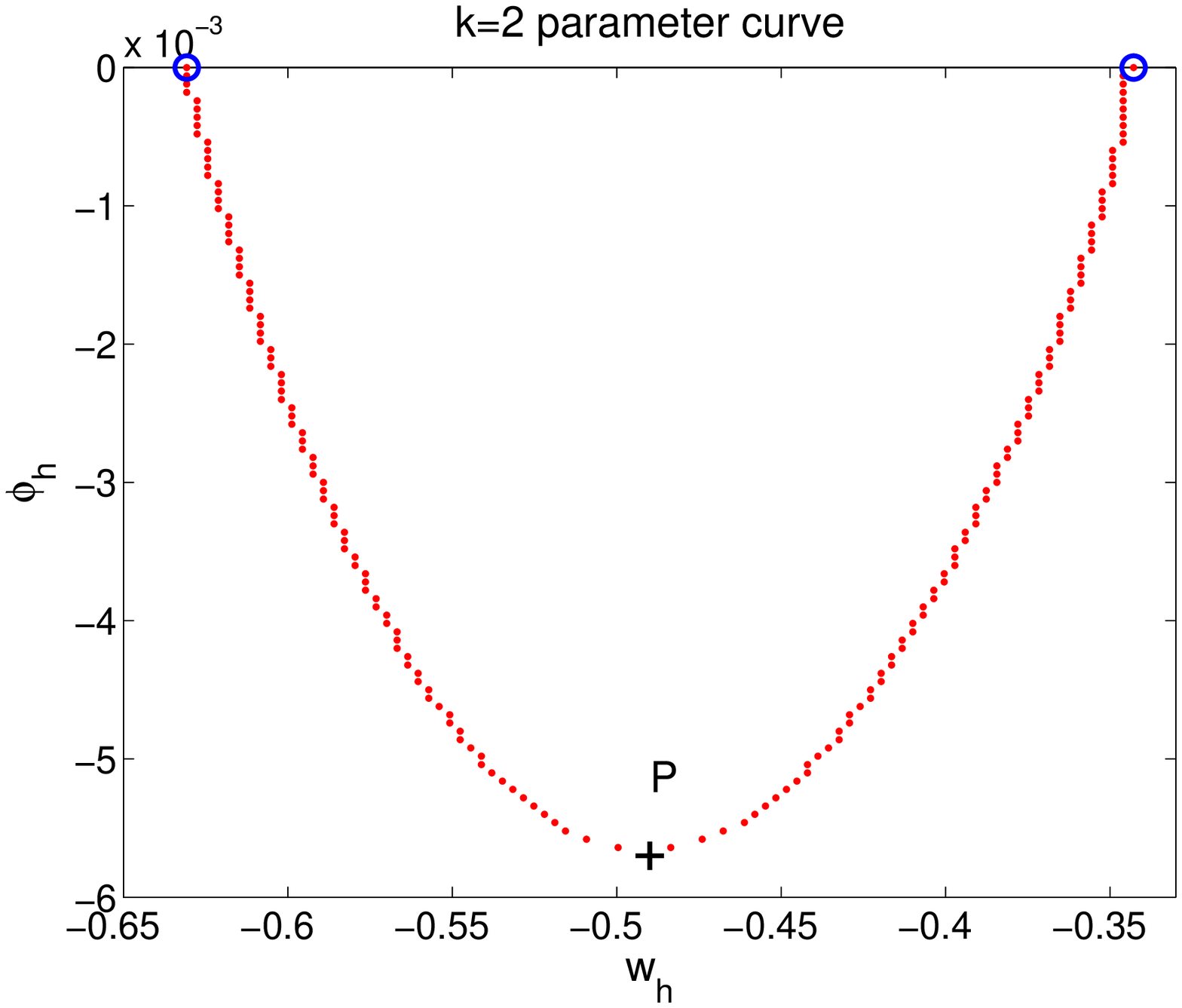}
\caption{\label{bhparsol1} The parameter curves for $k=1$ and $k=2$
solutions. The blue circles are the $w_h(1)$ and $w_h(2)$ of EYM
black hole solutions. }
\end{figure}

\begin{figure}[htp!]
\includegraphics[scale=0.4]{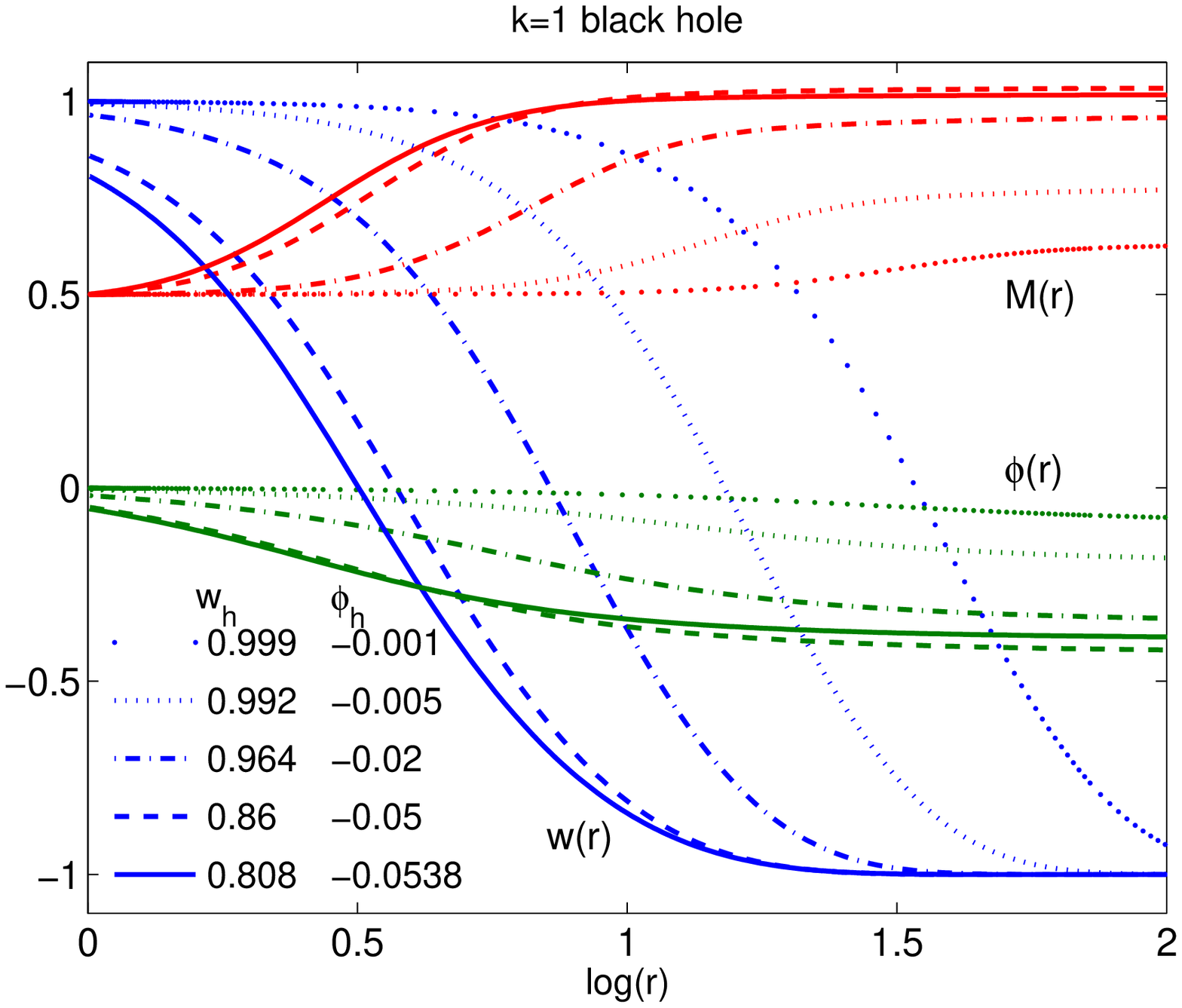}
\includegraphics[scale=0.4]{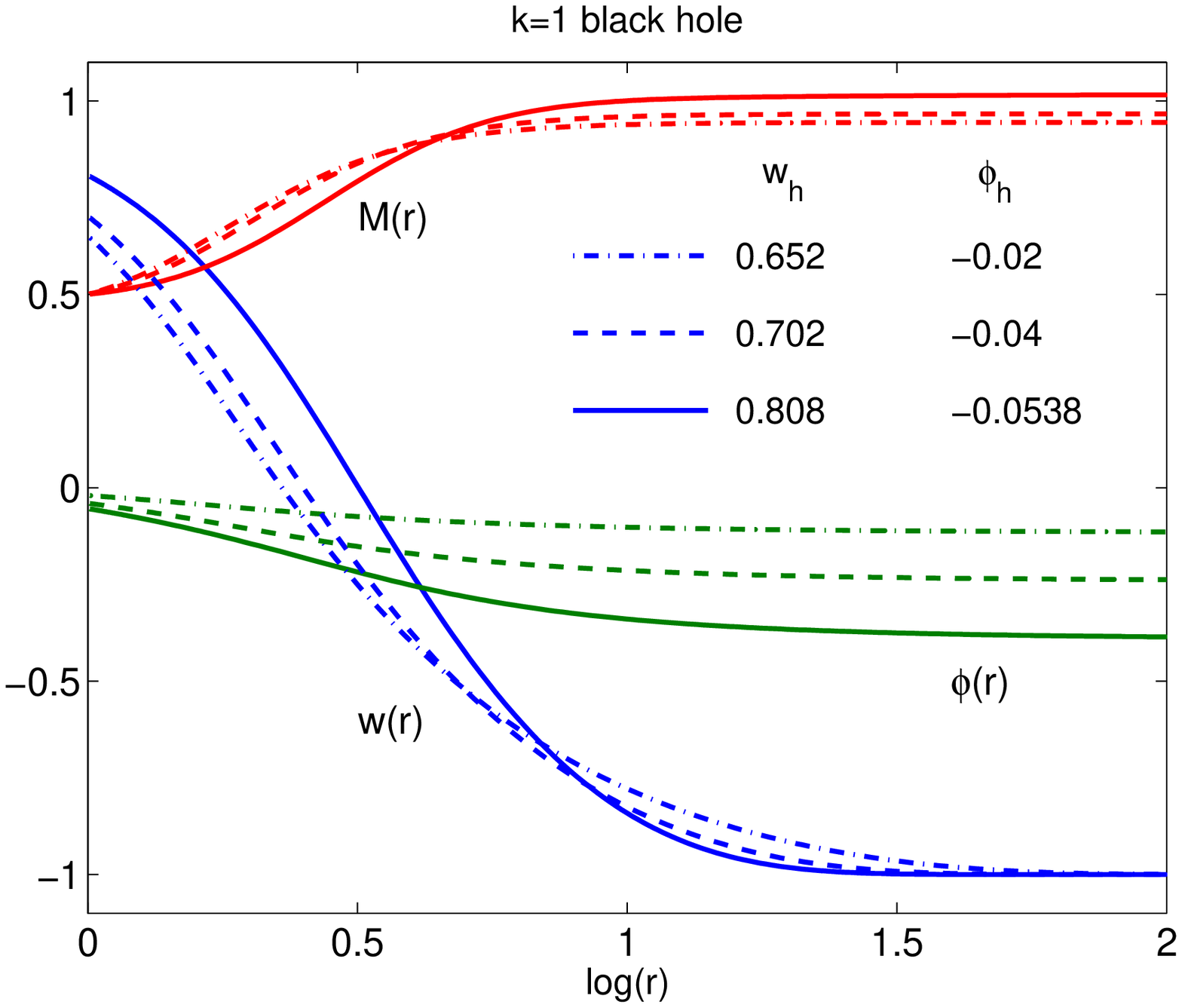}\\
\includegraphics[scale=0.4]{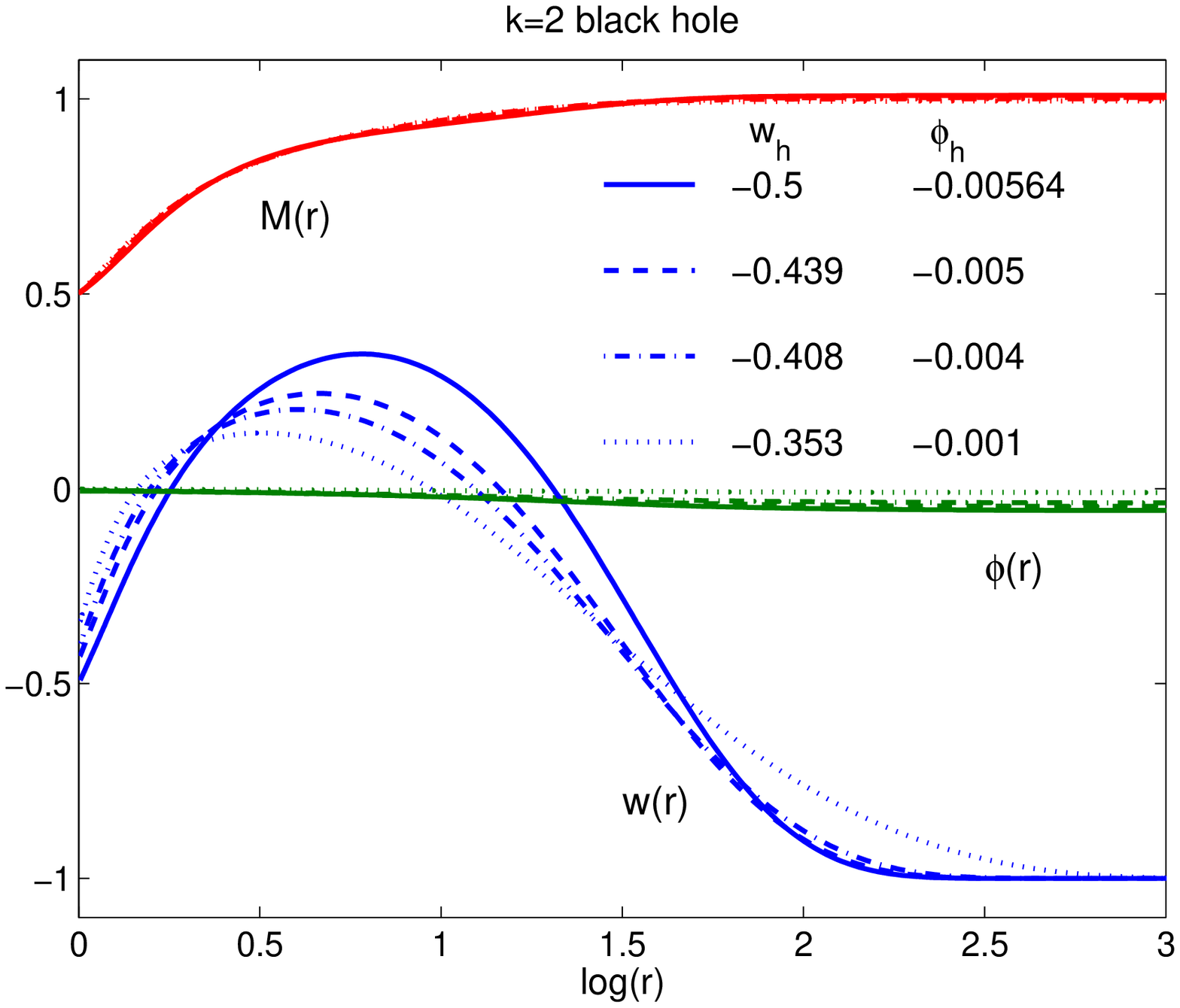}
\includegraphics[scale=0.4]{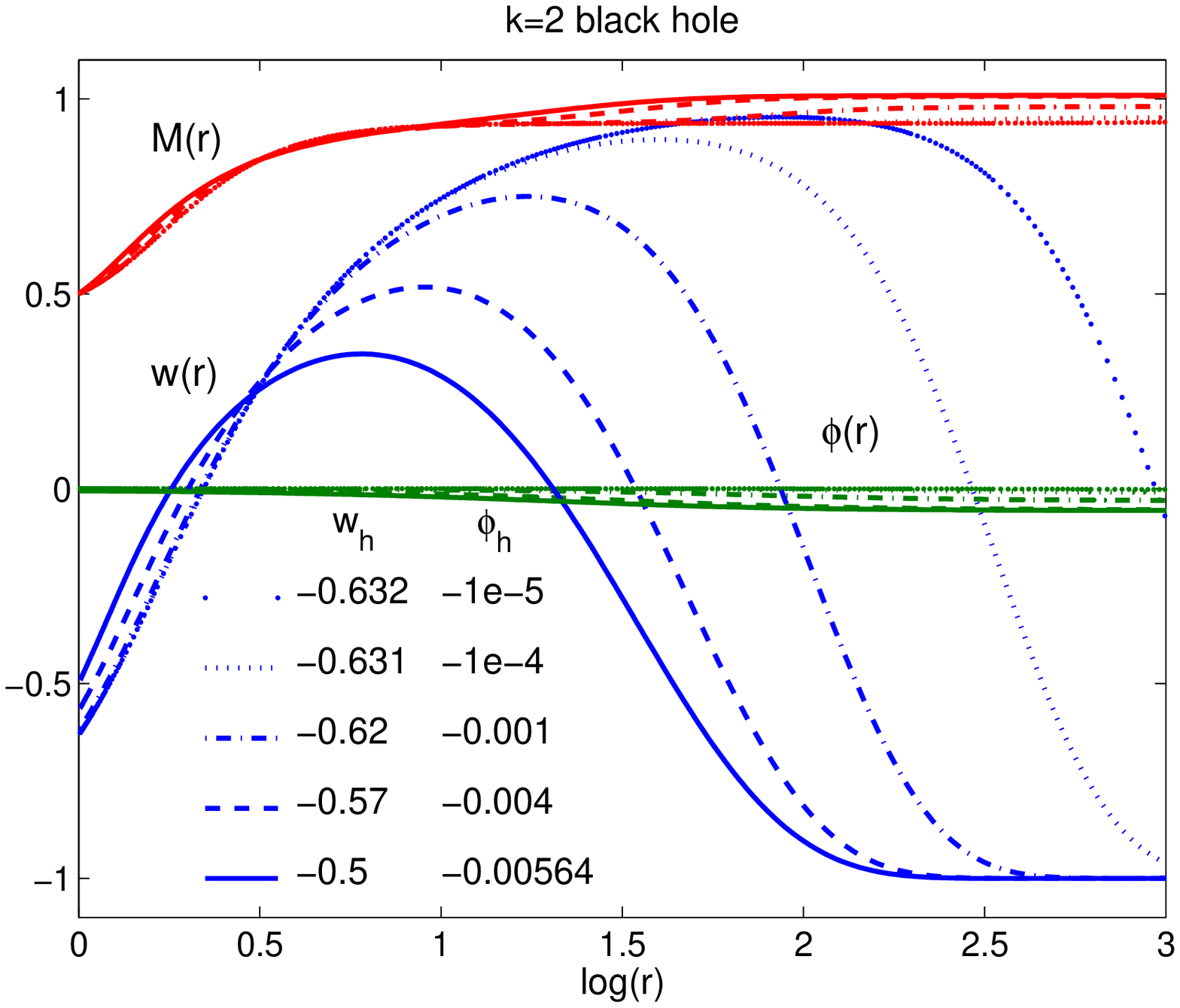}
\caption{\label{bhparsol12} The $k=1$ and $k=2$ black hole
solutions. The solid curves are solutions corresponding to peaks $P$
in the parameter curves. }
\end{figure}

Here we solve numerically for the boundary
conditions \refer{regasy1} and \refer{bhbc} using
$w_h$ and $\phi_h$ as shooting parameters and letting $r_h=1$. We
evaluate the boundary data at $r=r_h+10^{-2}$ first and then
integrate using $10^{-12}$ as tolerance towards $r=\infty$ to match
asymptotics \refer{regasy1}. The solutions to these conditions only
have kinetic energy contribution to the total
mass from the Higgs field. Again, the solutions exist and bifurcate along continuous
curves in the parameter space spanned by $w_h$ and $\phi_h$ (see
Fig. \ref{bhparsol1} for parameter curves and Fig. \ref{bhparsol12}
for solutions). The end points of these curves form a discrete set
$\{(w_h(k),\phi_h=0)\}$ where $k$ is the number of nodes of $w(r)$.
The solutions corresponding to these end points indeed are just the
black hole solutions of EYM theory found in Ref. \cite{pbizon} and a
list of the first five $w_h(k)$ could also be found there. For the
points $(w_h,\phi_h)$ on the curve that satisfy
$w_h(k)<w_h<w_h(k-1)$, the corresponding solutions of $w(r)$ have
$k$ nodes. Similar to the regular solutions, the black hole
solutions for each $k$ in general also bifurcate into two classes:
the $k$ node solutions that resemble the $k-1$ node black hole
solutions of EYM theory as $|w_h|$ increases along the parameter
curve and the $k$ node solutions that resemble the $k$ node black
hole solutions of EYM theory as $|w_h|$ decreases. Again we take the
bifurcation point to be the peaks of the parameter curves, where the
solutions have asymptotic masses that are relatively larger than
solutions with smaller $|\phi_h|$. This bifurcation behavior is
again believed to be due to the inner structure of the field
equations; but unlike the regular solutions case where we have a
simple relations \refer{reg1} and \refer{reg2} between the shooting
parameters $a$, $b_{1,0}$ and $\phi'(0)$ and $w'(0)$, the relation
\refer{bhbc} between $w_h$, $\phi_h$ and $w'(r_h)$, $\phi'(r_h)$ are
quite complicated. We therefore would not study this bifurcation in
more details, but just to remind the readers that the solutions here
are the black hole solutions to the EYMH theory with scalar field
minimally coupled to gauge field and gravity.

\section{Discussion}\label{secdis}

An important global property of these solutions is their charge.  The
non-Abelian $su(2)$ electrical charge $Q_E$ and magnetic
charge $Q_M$ of the gauge fields can be
defined as \cite{bh} \be \lb \ba{c}Q_E\\
Q_M\ea\rb=\frac{\tilde{g}}{4\pi}\int\dd S_{i0}\sqrt{-g}\lb \ba{c} F^{i0} \\
^*F^{i0}\ea\rb . \ee Using the given ans\"{a}tze and the
asymptotic values, it is found that for all type of solutions \be
Q_E=0,~Q_M\propto(1-w(\infty)^2)\tau_r =0, \ee  which means the
solutions are chargeless in the gauge corresponding to ans\"{a}tz
\refer{anzgauge}.

Another important issue is the stability of these solutions. For the
equation systems with conditions \refer{infsys1}, we know that both
their regular and black hole solutions are unstable with respect to
linear perturbations \cite{bbmsv,wm}. Because the stability depends crucially on the boundary conditions and asymptotics, we need to do a separate examination of the stabilities of the solutions found in this paper. We can carry out the perturbation of $SU(2)$ gauge fields in \refer{sphg} and Higgs field
in \refer{anzhiggs} along the line of Ref. \cite{bbmsv} and Ref. \cite{wm}
(note we have a sign difference in the definition of $w(r)$ compared to Refs. \cite{bbmsv,wm}):
\be
b(r,t)=-w'(r)z(r)e^{i\omega t}, ~d(r,t)=[w(r)^2-1]z(r)e^{i\omega
t},~\psi(r,t)=\frac{-w(r)-1}{2}\phi(r)z(r)e^{i\omega t}.\label{pertb}\ee  For regular solutions, we set $z(r)\equiv 1$. For black hole solutions, we choose $z(r)$ to be a real function that will be determined later.

Using this perturbation, it is found
that the perturbation equation take the same form \be H\Psi=-A\ddot{\Psi}
\ee where $\Psi\equiv(b(r,t),d(r,t),\psi(r,t))^T$ and $H$ and $A$
are matrix operators as in
Refs. \cite{bbmsv,wm}.
The expression for the eigenvalue square \be
\omega^2=\frac{\langle \Psi|H|\Psi\rangle}{\langle
\Psi|A|\Psi\rangle}, \ee is also the same as in Refs. \cite{bbmsv,wm}
\bea \langle \Psi|A|\Psi\rangle &=&\int_{r_0}^{\infty} \lsb \frac{r(w')^2}{S}+2\frac{(w^2-1)^2}{NS}+\frac{(w+1)^2\phi^2r^2}{4NS}\rsb z(r)^2 \dd r \label{papden}\\
\langle \Psi|H|\Psi\rangle &=&-\int_{r_0}^{\infty} \lsb2N(w')^2+2\frac{(w^2-1)^2}{r^2}+\frac{\phi^2}{2}(w+1)^2\rsb S\dd r \nonumber\\
&&+\int_{r_0}^{\infty} \lsb2N(w')^2+2\frac{(w^2-1)^2}{r^2}+\frac{\phi^2}{2}(w+1)^2\rsb \lb 1-z(r)^2\rb S\dd r\nonumber\\
&&+\int_{r_0}^{\infty} \lsb 2(w^2-1)^2+\frac{1}{4}(w+1)^2r^2\phi^2\rsb \lb z'\rb^2SN\dd r\nonumber\\
&&-\lsb 2(w^2-1)^2+\frac{1}{4}(w+1)^2r^2\phi^2\rsb SN zz'\longvert_{r=r_0}^{r=\infty}~,\label{phpnum}\eea where $N=1-2M/r$ and $S=(1-2M/r)^{-1}T^{-1}$, $r_0=0$ for regular solutions and $S=e^{-\delta}$, $r_0=r_h$ for black hole solutions. The terms in the last line of
\refer{phpnum} are the boundary terms that were only implicitly mentioned in Ref. \cite{wm}. Because the expressions \refer{papden} and \refer{phpnum} are exactly the same as in Refs. \cite{bbmsv,wm}, our arguments will follow exactly these references. We only need to pay attention to the applicability of our asymptotics \refer{regasy1}, \refer{regasy2} and \refer{bhasy1} and \refer{bhasy2}, because they are the only relevant difference from those of Refs. \cite{bbmsv,wm}. Below, we show the details of the arguments.

For regular solutions ($z(r)\equiv1$), it is clear that the last three lines in \refer{phpnum} drop and therefore $\langle \Psi|H|\Psi\rangle$ is clearly negative definite. From the asymptotics \refer{regasy1} and \refer{regasy2}, with little effort one can see that as $r$ becomes large all terms in the integrand of $\langle \Psi|A|\Psi\rangle$ vanish at least at the speed of $1/r^2$ and therefore $\langle \Psi|A|\Psi\rangle$ is finite.  This shows that the eigenvalue is negative and therefore the solutions are unstable.

For black hole solutions, we will show that for some properly chosen $z(r)$, $\langle \Psi|H|\Psi\rangle$ is negative definite and terms in the integrand of $\langle \Psi|A|\Psi\rangle$ decrease fast enough so that this term is finite. Moreover, we also need $z(r=r_h)=0$ for the perturbation \refer{pertb} to vanish at the horizon. These conditions can be satisfied by a $z(r)$ chosen according to Ref. \cite{wm,Volkov:1994dq}. This involves defining the tortoise coordinate $r^*$ by
\be \frac{\dd r^*}{\dd r}=\frac{1}{NS}\ee and a set of functions $z_k(r^*)$ by
\be z_k(r^*)=u\lb\frac{r^*}{k}\rb,~k\geq 1\ee where \bea  u(r^*)&=&u(-r^*),\nonumber\\
u(r^*)&=&1\mbox{ for }r^*\in[0,a],\nonumber\\
-D\leq \frac{\dd u(r^*)}{\dd r^*}&<&0\mbox{ for }r^*\in[a,a+1],\nonumber\\
u(r^*)&=&0\mbox{ for }r^*\in[a+1,\infty]. \eea
If we let $z(r)=z_k(r^*(r))$ in \refer{phpnum}, we see that the boundary terms (last line) vanish at horizon (corresponds to $r^*=-\infty$) because $z_k(r^*=\infty)=0$ and vanish at space infinity because of the asymptotics \refer{regasy1}, \refer{regasy2} and \refer{bhasy1} and \refer{bhasy2}. One can also see that the second and third lines of \refer{phpnum} will vanish uniformly as $k\to\infty$. This establishes that $\langle \Psi|H|\Psi\rangle$ is negative definite. This choice of $z(r)$ with $|z(r)|\leq1$ and the asymptotics \refer{regasy1}, \refer{regasy2} and \refer{bhasy1} and \refer{bhasy2} also guarantee that the integrand of \refer{papden} vanishes at least at the speed of $1/r^2$ as $r$ becomes large and therefore $\langle \Psi|A|\Psi\rangle$ is finite. Therefore again the eigenvalue is negative and the solutions are unstable.

Even though the EYM theory and its extensions have very interesting theoretical features, they are only relevant in few known objects in nature. One of these is the neutron stars, where the matter becomes very dense so that the gravity is very strong and the gluon fields become the fundamental degrees of freedom. The matter in neutron stars is usually described by effective Quantum Chromodynamical theories and chemical potential in this matter is large and important. Our original hope for this paper was to study the effect of a nonzero chemical potential to the EYMH theory. However, the reduced spherical symmetric form of the gauge fields -- ansatz \refer{anzgauge} -- that is used by previous studies on EYMH theory, dose not allow a consistent and simple introduction of the chemical potential that respects the spherical symmetry. Because a non-spherically symmetric setup in this case is very difficult, one can only attempt to construct a consistent theory with nonzero chemical potential and spherical symmetry from other forms of reduction of the most general gauge field ansatz \refer{sphg} and its gauge equivalents (e.g., see Ref. \cite{bh}). The result of this attempt will have to be reported later.

\vspace{1cm}
\noindent \textbf{Acknowledgement}\\
The author would like to thank V.A. Miransky, Alex Buchel and E.V.
Gorbar for helpful discussions. He is also thankful to a knowledgeable and encouraging referee for his/her valuable comments and recommendations. This work is supported by Natural
Sciences and Engineering Research Council of Canada and Ontario
Graduate Scholarship.

\end{document}